\documentclass[pre,twocolumn,floatfix,showpacs,showkeys]{revtex4}
\usepackage{amsmath}
\usepackage{amssymb}
\usepackage{amsfonts}
\usepackage{graphicx}
\usepackage{aecompl}
\usepackage{ae}
\usepackage{color}

\newcommand{\Sec}[1]{\ref{sec:#1}}

\newcommand{\Fig}[1]{Fig.~\ref{fig:#1}}
\newcommand{\Tab}[1]{Tab.~\ref{tab:#1}}

\def\bmath#1{\mbox{\boldmath$#1$}}

\begin{document}

\title{
Critical packing in granular shear bands
}

\author{
S. Fazekas$^{1,2}$, J. T\"or\"ok$^{1}$ and
J. Kert\'esz$^{1}$
}

\affiliation{
$^1$Department of Theoretical Physics,\\
$^2$Theoretical Solid State Research Group
of the Hungarian Academy of Sciences,\\
Budapest University of Technology and Economics (BME),
H-1111 Budapest, Hungary
}

\date{October 13, 2006}

\begin{abstract}
In a realistic three-dimensional setup, we simulate the slow deformation of
idealized granular media composed of spheres undergoing an axisymmetric
triaxial shear test. We follow the self-organization of the spontaneous
strain localization process leading to a shear band and demonstrate the
existence of a critical packing density inside this failure zone. The
asymptotic criticality arising from the dynamic equilibrium of dilation and
compaction is found to be restricted to the shear band, while the density
outside of it keeps the memory of the initial packing. The critical density
of the shear band depends on friction (and grain geometry) and in the limit
of infinite friction it defines a specific packing state, namely the
\emph{dynamic random loose packing}.
\end{abstract}



\maketitle

\section{Introduction}

Packing density of different particulate systems is of main interest for
scientific fields including, but not limited to, suspensions, metallic
glasses, molecular systems, and granular materials. In three dimensions,
for identical spheres, the face-centered cubic (FCC) packing is the maximum
possible \cite{weitz-sci04}. This fills the space with a volume fraction of
$\pi/(3\sqrt{2})\,\approx\,0.74$. Random arrangements have much lower
densities \cite{aste-pre05}. Different experiments and computer simulations
revealed that the largest obtainable volume fraction of a random packing of
identical spheres is around $0.64$. This is known as the \emph{random close
packing} (RCP) limit. A mathematical definition of this limit can be given
through the concept of \emph{maximally random jammed} state
\cite{torquato-prl00, ohern-pre03}.

Reaching the RCP limit needs careful preparation (e.g. tapping and
compression). If glass or marble beads are simply poured into a container
the volume fraction is usually only around $0.6$. A \emph{random loose
packing} (RLP) at its limit of mechanical stability obtained by immersing
spheres in a fluid and letting them settle has a volume fraction of $0.555$
\cite{onoda-prl90}. The volume fraction of RLPs obtained with different
methods (both experimental and numerical) show that this packing state is
less well defined than the RCP limit. Attempts made in order to relate RLP
to rigidity percolation \cite{onoda-prl90} and to critical density at
jamming of an assembly of (infinitely) rough spheres \cite{zhang-pre05},
are to be mentioned.

Already in 1885, Reynolds noted that dense granular samples dilate during
slow deformation \cite{reynolds-1885}. On the other hand, it is well known
that loose granular materials densify in such a process \cite{youd-jsmfd72,
torok-prl00}. Under slow shear the strain is usually localized to narrow
domains called shear bands. As it was first suggested by Casagrande
\cite{casagrande-jbsce36}, it is tempting to assume that in these failure
zones the system self-organizes its packing density to a critical value
independent of the initial packing state of the material.

While this hypothesis forms the basis of many continuum constitutive models
of soil mechanics since decades \cite{schofield-book68}, a general
micromechanical theory of shear band formation and of the involved
criticality is still missing. Progress, needed in order to deepen our
understanding of the critical state in shear bands, can be expected from
the remarkable development of experimental techniques (including Computer
Tomography \cite{desrues-geo96, desrues-ct04} and measurements in
microgravity \cite{alshibli-gtj00, batiste-gtj04}) and of simulation
techniques which become increasingly efficient as computational power grows
\cite{poeschel-05}.

Shearing of granular materials has been investigated in many different
geometries and specially designed laboratory tests (for recent results see
\cite{desrues-ct04, batiste-gtj04}). Such experimental studies revealed
complex localization patterns and presented evidence for the existence of a
critical particle density inside the shear bands. The importance of
computer simulations is enhanced by the fact that they make possible
studies which are difficult to control in experiments (e.g. friction
dependence) and they facilitate the measurement of hardly accessible
quantities (e.g. volume fraction inside the shear bands).

The critical density, in numerical studies, is often studied only in
special conditions when shearing extends to the whole volume of the
samples. This allowed for discussing the criticality based simply on global
behavior (e.g.\ dilatancy). Without reference to shear bands, many
qualitative effects were already pointed out in both mathematical models
\cite{piccioni-pre00, rothenburg-ijss04} and simulations
\cite{zhuang-jcp95, rothenburg-ijss04}. However, such studies neglected the
involved localization phenomena inevitable in real situations and
disregarded the self-organizing manner in which the packing state of the
shear bands is usually formed.

A principal parameter which controls the dynamic equilibrium between
dilation and compaction in fully developed shear bands is the friction
between the grains. Intuitively, a system of frictionless grains can be
sheared at a large packing density (close to the RCP limit) because the
grains (under slow shear) can easily rearrange in compact configurations.
At large friction the rearrangement of the grains is hindered by friction,
consequently the packing density of the shear bands is expected to define a
low density state close to the RLP limit.

The aim of this Paper is to study the emergence of a critical packing state
in sheared granular media and to present its relation to shear bands as
well as its dependence on friction.

\section{Simulations}

\begin{figure}[t!]
\includegraphics{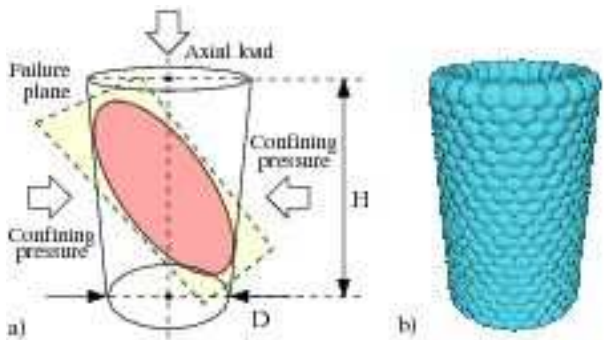}
\caption{
(Color online)
a) Grains placed between two horizontal platens and surrounded by an
elastic membrane were subjected to a vertical load and a lateral confining
pressure. b) The membrane was modeled with overlapping spheres
\cite{fazekas-pre06}.
}
\label{fig:simmod}
\end{figure}

We investigate numerically an axisymmetric triaxial shear test (see
\Fig{simmod}). This consists of the slow compression of a cylindrical
sample enclosed between two end platens. The sample is surrounded by an
elastic membrane on which an external confining pressure is applied. The
end platens are pressed against each other in a strain controlled way. The
upper platen is allowed to tilt. In certain conditions a planar shear band
is formed \cite{desrues-ct04}. Using different initial packing densities
and friction properties of the grains as well as identifying the grains in
the failure zones makes it possible to study the critical packing density
inside the shear bands.

The simulations, which are going to be presented here, are based on a
standard three-dimensional Distinct Element Method (DEM)
\cite{cundall-geo79}. We implemented the Hertz contact model
\cite{landau-book70} with appropriate damping combined with a frictional
spring-dashpot model \cite{poeschel-05}. The normal $F_n$ and the
tangential ${\bf F}_t$ forces are calculated as
\begin{eqnarray}
  F_n &=&
    {\kappa}_n {\delta}_n^{3/2} - \gamma_n {\delta}_n^{1/2} {v_n} \\
  {\bf F}_t &=&
    {\kappa}_t {\bmath{\delta}}_t - \gamma_t {\bf v}_t
\end{eqnarray}
where ${\kappa}_{n} \,{=}\, 10^{6}\,\mathrm{N/m}^{3/2}$,
${\kappa}_{t}\,{=}\, 10^{4}\,\mathrm{N/m}$, ${\gamma}_{n} \,{=}\,
1\,\mathrm{N\,s/m}^{3/2}$, and ${\gamma}_{t} \,{=}\, 1\,\mathrm{N\,s/m}$
are normal and tangential stiffness and damping coefficients, ${\delta}_n$
and ${\bmath{\delta}}_t$ are normal and tangential displacements, and
${v}_n$ and ${\bf v}_t$ are the normal and tangential relative velocities.

The numerical values are chosen to realize the hardest material that we
could safely simulate with the minimal damping preserving the numerical
stability of the calculations. With the above stiffness and damping
coefficients, the inverse of the average eigenfrequency of contacts, in
both normal and tangential direction, is more than one order of magnitude
larger than the used integration time step $\Delta t\,=\,10^{-6}\,s$. This
assured that the noise level induced by numerical errors is kept low.

With relatively small samples (made up of $27000$ particles) but in a
realistic geometry we have succeeded to reproduce shear band morphologies
\cite{fazekas-pg05, fazekas-pre06} known from experiments
\cite{desrues-ct04, batiste-gtj04}. In order to study the criticality of
these shear bands, we prepared homogeneous initial configurations of
different volume fractions using the deposition method described in
\cite{fazekas-pre06}.

We used a particle distribution similar to those encountered in
experimental studies of idealized granular materials. Our particles are
spherical, they have equal mass density ($2.5\cdot 10^3~\mathrm{kg/m}^3$),
equal friction coefficient, and their diameters are set according to a
narrow Gaussian distribution with mean $d = 0.9~\mathrm{mm}$ and standard
deviation of $2.77\%$. The prepared cylindrical samples, having diameter $D
= 23.3\ d$, consisted of $20000$ to $27000$ spherical grains as required
by a prescribed packing density and the $H \approx 2.2\ D$ geometrical
constraint, where $H$ is the height of the samples.

Initially the particles were placed randomly in a tall cylinder (about 3
times taller than $H$). They were given small downwards velocities in such
a way that they all collided approximately at the same time. The upper
platen was pressed on top of the packing to hold it together. This method
provides an efficient way to produce a homogeneous random packing. The
volume fraction of the prepared samples could be controlled in the full RLP
to RCP range (see \Tab{volfrac}) by varying the coefficient of friction
$\mu_0$ which was applied during this phase.

\begin{table}[t!]
\begin{center}
\begin{tabular}{c|ccccccc}
$\mu_0$ && 0.8   & 0.5   & 0.3   & 0.2   & 0.1   & 0.0   \\
\hline
$\eta_0$   && 0.555 & 0.562 & 0.578 & 0.599 & 0.621 & 0.641 \\
\end{tabular}
\end{center}
\caption{
Volume fraction $\eta_0$ of samples prepared with different coefficients of
friction $\mu_0$. With each $\mu_0$ we prepared $2$ samples having the same
$\eta_0$ within $0.2\%$ relative error.
}
\label{tab:volfrac}
\end{table}

After preparation, the friction coefficient of the particles was set to a
new value $\mu$ independent of $\mu_0$. During the simulations, similarly
to \cite{fazekas-pre06}, we compressed the samples vertically at zero
gravity and $0.5\,\mathrm{kPa}$ confining pressure. The bottom platen was
fixed. The upper platen moved downward with a constant velocity, inducing
an axial strain rate of $20\,\mathrm{mm/s}$. During compression the upper
platen could freely tilt along any horizontal axis with rotational inertia
$I\,{=}\,10^{-7}\,kg\,m^2$.

The lateral membrane surrounding the sample was modeled with approximately
$15000$ identical, overlapping, non-rotating, frictional spheres connected
with elastic springs. The stiffness of the springs was set to
$\kappa_s\,{=}\,0.5\,\mathrm{N/m}$. This prevented the particles from
escaping by passing through the membrane. The membrane particles were
initially arranged in a triangular lattice (\Fig{simmod} (b)). The
confining pressure was applied on the triangular facets formed by the
neighboring ``membrane nodes'' as described in \cite{fazekas-pre06} (see
also \cite{tsunekawa-pg01, sakaguchi-eg01, cui-pg05}).

It is worth mentioning that Cui and O'Sullivan \cite{cui-pg05} have introduced
a technique which speeds up calculations by computing only a section of the
cylindrical sample. This allows for larger samples but requires that the
symmetry of the system is kept during compression, and thus eliminates the
possibility of symmetry breaking strain localization, which arises
spontaneously \cite{fazekas-pre06} if tilting of the upper platen is not
suppressed (see \Fig{simmod} (a)).

We have executed several simulation runs. The grain-platen and
grain-membrane contacts were calculated similarly to grain-grain contacts
including the friction properties. The samples of different initial volume
fractions (see \Tab{volfrac}) were first compressed using the same
coefficient of friction $\mu=0.5$. Later, we compressed the densest samples
($\eta_0=0.641$) with $10$ different friction coefficients $\mu \in
\{0,\:0.1, \dots 0.9\}$. For each set of parameters, two simulation runs
were executed using specimens prepared with different random seeds.

During compression, we measured \emph{locally} the shear intensity $S$ and
the volume fraction $\eta$. The regular triangulation of the spherical
grains \cite{lee-book91, edelsbrunner-alg96} was used to define these
quantities. The local volume fraction is given by the ratio of the volumes
of a grain and its regular Voronoi cell. The local shear intensity is
calculated from the macroscopic strain tensor derived from particle
displacements \cite{daudon-pg97, fazekas-pre06}. Using the eigenvalues
$\varepsilon_k$ of this tensor, we defined the local shear intensity as
\begin{equation}
S = \max_{k} \left| \varepsilon_k - \frac{1}{3} \sum_{l}
\varepsilon_l \right|.
\end{equation}
To overcome fluctuations due to random packing and rearrangements, we
calculated spatial averages up to 3rd order neighbors along the regular
triangulation.

\section{Identification of high shear intensity regions}

Strain localization in dense and loose samples shows substantial
differences \cite{desrues-ct04}. In dense samples shear bands are usually
formed after a short plastic deformation and inside them the local packing
density is lower than in the bulk (i.e. the regions outside of the shear
bands). Since denser parts are more stable, the position of the shear bands
remains unchanged for the whole duration of a shear test. Contrary, in
loose samples the shear bands have a slightly higher packing density than
the bulk and hence the position of the shear bands is likely to change and
to move around the whole sample. This leads to more or less homogeneous
samples with local packing densities close to the packing density of the
shear bands.

\begin{figure}[t!]
\includegraphics{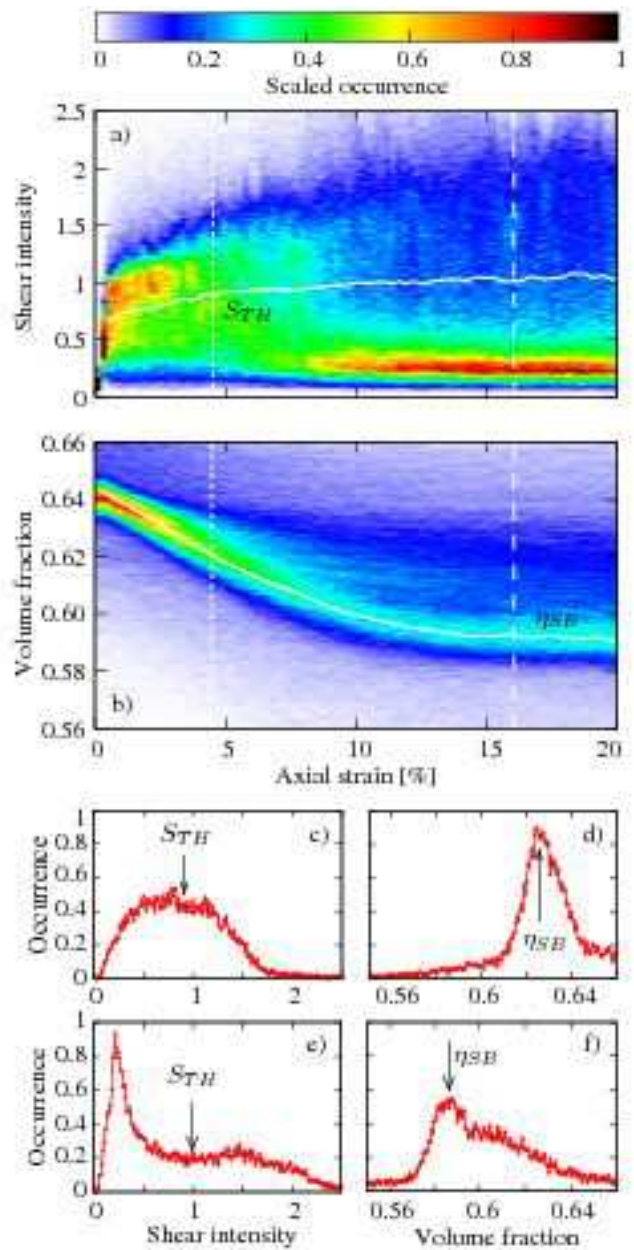}
\caption{
(Color online)
Example of shear intensity and volume fraction histogram maps
($\eta_0=0.641$, $\mu=0.5$). The shear intensity is measured in arbitrary
units. On (a, b) the occurrences (scaled to $[0,1]$) are encoded with the
color scale shown at the top. On (a) the white curve marks the shear
intensity threshold $S_{TH}$. On (b) it marks the shear band volume
fraction $\eta_{SB}$. The dotted and dashed vertical lines on (a, b) at
$4.4\%$ and $16\%$ axial strain mark the position of the histograms (c, d)
and (e, f), respectively.
}
\label{fig:maps}
\end{figure}

The general algorithmic identification of failure zones based on geometric
methods is difficult, especially, regarding the identification of the
non-persistent shear bands of the loose samples. Nevertheless, based on the
calculated local shear intensity, individual grains can be categorized to
be part of the failure zones or the bulk providing that a good enough
threshold separating these two classes can be found. A histogram technique
seems to be a perfect candidate for this.

\Fig{maps} presents histograms of the local shear intensity and local
volume fraction. It shows the main aspects of the strain localization
process observed in one of our simulations executed with a sample
having $\eta_0=0.641$ and $\mu=0.5$. At the beginning of the test, up
to approximately $6\%$ axial strain, in shear intensity histograms
(see \Fig{maps} (a, c)), we find a single peak at medium $S$. This means
that almost all particles rearrange simultaneously and consequently
the sample experiences a more or less plastic deformation. This is
underlined by the fact that the local volume fraction has just one
strong peak (\Fig{maps} (d)), indicating that the sample is still
homogeneous.

At higher ($>6\%$) axial strain a shear band is formed. This is
localized to a planar failure zone of width of approximately $10$
particle diameters and is characterized by much higher $S$ than the
bulk. In the bulk the shear intensity fluctuations are small, while
these fluctuations are large in the shear band. Consequently, in shear
intensity histograms (\Fig{maps} (e)), we find a narrow peak at low
$S$, corresponding to the bulk, and a wide peak at high $S$,
corresponding to the shear band. The volume fraction histogram
does also become more structured showing evidence of a non-homogeneous
material. The narrow peak at low volume fraction corresponds to the
shear band while the bulk produces a much wider distribution at a
higher volume fraction (\Fig{maps} (f)).

Motivated by this separation, we computed a shear intensity threshold
$S_{TH}$, which could be used to define two classes of shear intensity
values (low and high) and to classify the grains accordingly into shear
band and bulk. For this we used Otsu's threshold selection method
\cite{otsu-tsmc79} described in Appendix \Sec{appendix_otsu}. This
histogram technique minimizes the within-class variance and maximizes the
separation of classes, and thus gives an ideal solution to our problem. We
have also tested another threshold selection method modeling the histograms
with the sum of two Gaussian functions, however, numerically this proved to
be less stable and less reliable.

The condition $S > S_{TH}$, made it possible to identify the grains in high
shear intensity regions -- where shear bands emerge -- and thus the average
volume fraction $\eta_{SB}$ of these regions could be calculated. The local
packing density in shear bands is found to have small fluctuations and to
give a peak in the volume fraction histograms. This coincides with
$\eta_{SB}$ (see \Fig{maps} (b, f)), giving a self validation of the
method.

Let us note that even if \Fig{maps} (e) does not suggest a
\emph{clean} separation of the grains into those within and those outside
the shear bands -- i.e. there is no shear intensity gap between the two
regimes -- this is not crucial for $\eta_{SB}$. Adding an artificial random
noise of $10\%$ to $S_{TH}$ does influence the resulting $\eta_{SB}$ only
within $0.5\%$.

We also mention here, that before strain localization takes place the
shear intensity histograms have only one peak (see \Fig{maps} (c)). In this
case, the threshold given by Otsu's method, which falls on the middle of
the peak, is not physically relevant. However, as the sample is
homogeneous, the selected samples in high shear intensity regions still
give the volume fraction which is close to the average volume fraction of
the whole sample. This can be verified on \Fig{maps} (d).

\section{Results}

\begin{figure}[t!]
\includegraphics{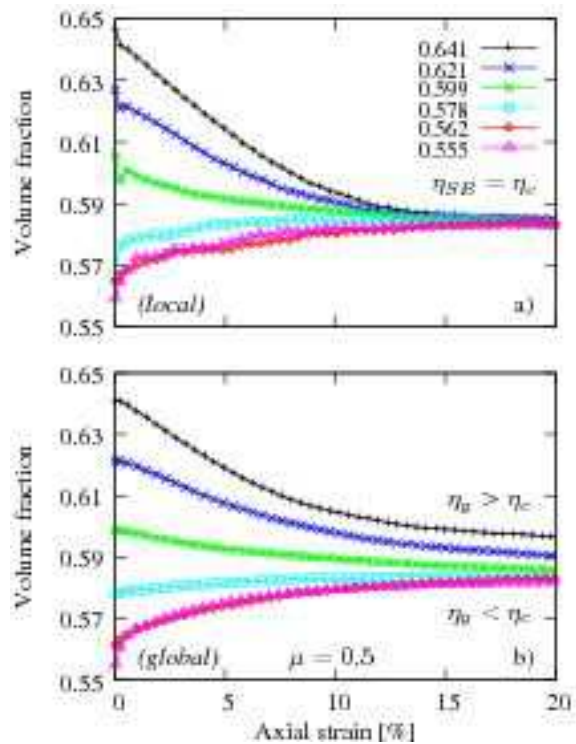}
\caption{
(Color online)
Volume fraction measured in high shear intensity regions (a) and
globally (b) as function of the axial strain. The different lines
correspond to different initial packing densities $\eta_0$
(see \Tab{volfrac}) decreasing from top to bottom.
The two panels use the same notations.
}
\label{fig:volfrshbglob}
\end{figure}

As expected \cite{desrues-ct04, batiste-gtj04}, we found that due to strain
localization at the end of the shear tests the global volume fraction of
the samples is not equal to the packing density of the high shear intensity
regions and thus global measurements cannot be used to characterize the
properties of failure zones. The behavior of both dense and loose samples
demonstrates that in the shear bands, the initial packing conditions
are canceled and a critical volume fraction $\eta_c$ is reached in a
self-organizing manner independently of the initial density of the
tested granular specimens (\Fig{volfrshbglob} (a)).

The criticality is found to be restricted to the shear bands. The global
volume fraction $\eta_g$ calculated from the total volume of the samples
does not converge to $\eta_c$ (\Fig{volfrshbglob} (b)). This behavior is
expected to be more pronounced on larger systems. The dense samples are
characterized by $\eta_g > \eta_c$. This demonstrates that the dilatancy
\cite{reynolds-1885} is concentrated to the shear bands. Contrary, for
loose samples $\eta_g < \eta_c$, however, the specimen is only slightly
looser outside the shear bands. This gives a direct proof of shear induced
compaction \cite{youd-jsmfd72}.

\begin{figure}[t!]
\includegraphics{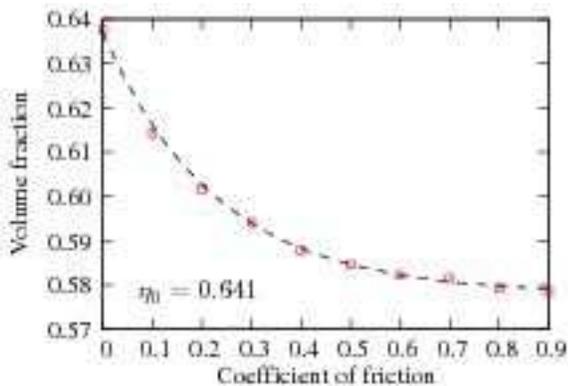}
\caption{
(Color online)
Critical packing of shear bands as function of friction
measured at $20\%$ axial strain. The fitted curve shows
$\eta_c(\mu)=\eta_c^\infty-(\eta_c^\infty-\eta_c^0)\exp(-\mu/\mu_c^0)$,
where $\eta_c^0=0.637$, $\eta_c^\infty=0.578$, and $\mu_c^0=0.23$.
}
\label{fig:volfrcrit}
\end{figure}

As we could see, at given friction, the critical packing of shear bands is
independent of the initial density. It is, however, a further question
whether it depends on friction. We studied quantitatively this effect for
samples with initial density $\eta_0\,{=}\,0.641$. At $\mu\,{=}\,0$ the
high shear intensity regions have a large volume fraction $\eta_c^0 \equiv
\eta_c(0) = 0.637 \pm 0.002$, which is only slightly smaller than $\eta_0$,
showing that frictionless granular systems can be sheared at densities very
close to the RCP limit.

In frictional systems, the volume fraction of the fully developed shear
bands is substantially lower than the initial volume fraction (see
\Fig{volfrcrit}). With increasing $\mu$, this decreases and converges to a
limit which we estimate to $\eta_c^\infty \equiv \lim_{\mu\to\infty}
\eta_c(\mu) = 0.578 \pm 0.003$ based on the exponential extrapolation of
our data.

This limit volume fraction depends only on geometry factors such as shape
and size distribution of the grains and is characteristic to the dynamic
equilibrium between dilation and compaction developed in a self-organized
manner through strain localization. Based on the value of $\eta_c^\infty$
and this latter aspect, the corresponding asymptotic state -- which we
refer to as the \emph{dynamic random loose packing} (DRLP) -- should be
distinguished from the static RLP limit.

\section{Conclusions and discussion}

We have presented distinct element simulations of axisymmetric triaxial
shear tests at zero gravity and low confining pressure. Due to spontaneous
strain localization, shear bands were formed. Using a histogram technique,
we identified the grains in high shear intensity regions, which at large
axial strains coincide with the shear bands. We measured the packing
density $\eta_{SB}$ inside these failure zones and we found that in fully
developed shear bands $\eta_{SB}$ approaches a critical value $\eta_c$
independent of the initial density of the samples. This in agreement with
Casagrande's \cite{casagrande-jbsce36} observation made for sandy soils
seven decades before and also with recent experiments \cite{desrues-geo96,
desrues-ct04, alshibli-gtj00, batiste-gtj04} and numerical studies
\cite{zhuang-jcp95, piccioni-pre00, rothenburg-ijss04}.

Rothenburg and Kruyt \cite{rothenburg-ijss04} obtained similar results in
two-dimensional simulations of biaxial shear tests. They have
presented a theory of the average coordination number of sheared granular
media and derived a law for its evolution during slow deformations.
Analyzing the relationship between volume fraction and average coordination
number, they conclude that a proper characterization of granular
media undergoing shear deformation should be based on packing density.

We have shown that the criticality is restricted to the shear bands and
global measurements (such as dilatancy) are unsuitable for the
investigation of the properties of sheared granular materials in realistic
situations, where strain localization is inevitable. To our knowledge, it is
the first time that the critical packing density of shear bands was
evidenced based on simulations of a realistic three-dimensional setup and
spontaneous strain localization revealing the self-organizing manner in
which the packing state of the shear bands is developed.

We have further shown that $\eta_c$ depends on the coefficient of friction
$\mu$ and in the limit $\mu \to \infty$ it converges to a value
$\eta_c^\infty$, which we have calculated within the accuracy of our
simulations. The found $\eta_c^\infty$ defines a low density \emph{dynamic
random loose packing} (DRLP) state, which is characteristic to the dynamic
equilibrium between dilation and compaction in the shear bands and depends
only on the geometry of the grains. Based on the underlying mechanism, we
argue that the asymptotic packing state of shear bands differs from the
static RLP limit.

This result should be also compared with the findings presented recently by
Zhang and Makse \cite{zhang-pre05} regarding the critical density of
granular materials at jamming transition. The importance of these findings
lie in the fact that jamming is a basic concept which through a unifying
phase diagram \cite{liu-nat98, trappe-nat01} connects granular matters with
a variety of other systems including dense particulate suspensions and
effects such as diverging viscosity at a maximum packing fraction
\cite{stickel-arfm05, ovarlez-jr06}.

In quasistatic limit, Zhang and Makse \cite{zhang-pre05} reported a
monotonous decrease of the critical packing density as function of
friction. For low friction, the density of the shear bands found in our
simulations is lower than at jamming found by Zhang and Makse
\cite{zhang-pre05}, while at high friction the situation is reversed. This
indicates a natural separation of low and high density regions with
possibly different mechanisms of dissolving the jammed state.

As a final remark let us note that our results are derived for idealized
granular materials composed of spheres having a narrow size distribution.
It is well known that for non-spherical grains and wide size distributions
the packing efficiency increases \cite{donev-sci04}, which should be also
reflected in the packing density of the shear bands. This could be the
reason why experimental results on sand reveal smaller volume fractions in
shear bands \cite{desrues-ct04, batiste-gtj04} than the values found in our
simulations.

\section{Acknowledgments}

This research was carried out within the framework of the ``Center for
Applied Mathematics and Computational Physics'' of the BME, and it was
supported by OTKA F047259 and T049403, and the P\'eter P\'azm\'any program
RET-06/2005. S.F. thanks D. Chetverikov for the introduction to Otsu's
method.

\begin{appendix}

\section{Otsu's threshold selection method}
\label{sec:appendix_otsu}

Otsu's method \cite{otsu-tsmc79} is a histogram technique known from
Digital Image Processing, where it is typically used to transform grayscale
images into two component (black and white) images.

Let us consider a normalized histogram $P(i)$, i.e. a histogram with the
property
\begin{equation}
  \sum_{i}\:P(i) = 1,
\end{equation}
where $i$ is the bin index. The mean
$\mu$ and the variance $\sigma^2$ can be calculated as
\begin{eqnarray}
  \mu      &=& \sum_{i}\: i \: P(i), \\
  \sigma^2 &=& \sum_{i}\: (i-\mu)^2 \: P(i).
\end{eqnarray}

Let us further consider a candidate threshold $t$ and split the histogram
in two parts ${\cal I}_1(t)=\{i\:|\:i \leq t\}$ and ${\cal
I}_2(t)=\{i\:|\:i>t\}$. With $k \in \{1,2\}$ and
\begin{equation}
  q_k(t)=\sum_{i \in {\cal I}_k(t)}\:P(i),
\end{equation}
the mean $\mu_k(t)$ and variance $\sigma_k^2(t)$ of the two
parts are defined by the equations
\begin{eqnarray}
  q_k(t)\:\mu_k(t)      &=&
    \sum_{i \in {\cal I}_k(t)}\: i \: P(i), \\
  q_k(t)\:\sigma_k^2(t) &=&
    \sum_{i \in {\cal I}_k(t)}\: \big(i-\mu_k(t)\big)^2 \: P(i).
\end{eqnarray}

The \emph{within-class variance}
\begin{equation}
  \sigma_W^2(t) = q_1(t)\sigma_1^2(t)+q_2(t)\sigma_2^2(t)
\end{equation}
is an inverse measure of the compactness of classes.
The \emph{between-class variance}
\begin{equation}
  \sigma_B^2(t) = q_1(t)q_2(t)\big(\mu_1(t)-\mu_2(t)\big)^2
\end{equation}
is a measure of the separation of classes. It is easy to show that
$\sigma_W^2(t)+\sigma_B^2(t)=\sigma^2$. Otsu \cite{otsu-tsmc79} proposed to
calculate an optimal threshold $t=t_{opt}$ by either minimizing
$\sigma_W^2(t)$ or maximizing $\sigma_B^2(t)$.

Maximizing $\sigma_B^2(t)$ is easier. It can be seen that
\begin{eqnarray}
  q_2(t) &=& 1-q_1(t),\\
  \mu_2(t) &=& \frac{\mu-q_1(t)\mu_1(t)}{q_2(t)},
\end{eqnarray}
and thus
\begin{equation}
  \sigma_B^2(t) = \frac{q_1(t)}{1-q_1(t)} \big(\mu_1(t)-\mu\big)^2.
\end{equation}

For each candidate threshold $t$,
$q_1(t)$ and $\mu_1(t)$ can be calculated with the recursive formula
\begin{eqnarray}
  q_1(t+1) &=& q_1(t)+P(t+1), \\
  \mu_1(t+1) &=& \frac{q_1(t)\:\mu_1(t)+(t+1)\:P(t+1)}{q_1(t+1)},
\end{eqnarray}
where $q_1(0)=P(0)$ and $\mu_1(0)=0$.

Both $q_1(t)$ and $\mu_1(t)$ are increasing monotonously with $t$,
consequently the maximum of $\sigma_B^2(t)$ is
well-defined, except for degenerated cases which must be handled
separately. The optimal threshold $t_{opt}$ is given by the smallest
candidate threshold $s$ which satisfies the equation
\begin{equation}
  \sigma_B^2(s) = \max_{t} \sigma_B^2(t).
\end{equation}
Because $\sigma_W^2(t_{opt})+\sigma_B^2(t_{opt})=\sigma^2$,
the method both minimizes the within-class variance and maximizes the
separation of classes.

\end{appendix}

\bibliography{packing}

\end{document}